\documentclass[12pt]{article}
\usepackage{amsfonts}
\usepackage{amsmath}

\advance \topmargin by -\headheight
\advance \topmargin by -\headsep
     
\evensidemargin \oddsidemargin
 \date{}    
\begin{document}

\topmargin -.6in
\def\rf#1{(\ref{eq:#1})}
\def\lab#1{\label{eq:#1}}
\def\nonu{\nonumber}
\def\br{\begin{eqnarray}}
\def\er{\end{eqnarray}}
\def\be{\begin{equation}}
\def\ee{\end{equation}}
\def\eq{\!\!\!\! &=& \!\!\!\! }
\def\ba{\be\begin{array}{c}}
\def\ea{\end{array}\ee}
\def\foot#1{\footnotemark\footnotetext{#1}}
\def\lb{\lbrack}
\def\rb{\rbrack}
\def\llangle{\left\langle}
\def\rrangle{\right\rangle}
\def\blangle{\Bigl\langle}
\def\brangle{\Bigr\rangle}
\def\llb{\left\lbrack}
\def\rrb{\right\rbrack}
\def\Blb{\Bigl\lbrack}
\def\Brb{\Bigr\rbrack}
\def\lcurl{\left\{}
\def\rcurl{\right\}}
\def\({\left(}
\def\){\right)}
\def\v{\vert}                     
\def\bv{\bigm\vert}               
\def\lskip{\vskip\baselineskip\vskip-\parskip\noindent}
\def\mskp{\par\vskip 0.3cm \par\noindent}
\def\sskp{\par\vskip 0.15cm \par\noindent}
\def\bc{\begin{center}}
\def\ec{\end{center}}

\def\tr{\mathop{\rm tr}}                  
\def\Tr{\mathop{\rm Tr}}                  
\makeatletter
\newcommand{\rd}{\@ifnextchar^{\DIfF}{\DIfF^{}}}
\def\DIfF^#1{%
   \mathop{\mathrm{\mathstrut d}}%
   \nolimits^{#1}\gobblespace}
\def\gobblespace{\futurelet\diffarg\opspace}
\def\opspace{%
   \let\DiffSpace\!%
   \ifx\diffarg(%
   \let\DiffSpace\relax
   \else
   \ifx\diffarg[%
   \let\DiffSpace\relax
   \else
   \ifx\diffarg\{%
   \let\DiffSpace\relax
   \fi\fi\fi\DiffSpace}
\newcommand{\deriv}[3][]{\frac{\rd^{#1}#2}{\rd #3^{#1}}}
\providecommand*{\dder}[3][]{%
\frac{\rd^{#1}#2}{\rd #3^{#1}}}
\providecommand*{\pder}[3][]{%
\frac{\partial^{#1}#2}{\partial #3^{#1}}}
\newcommand{\renewoperator}[3]{\renewcommand*{#1}{\mathop{#2}#3}}
\renewoperator{\Re}{\mathrm{Re}}{\nolimits}
\renewoperator{\Im}{\mathrm{Im}}{\nolimits}
\providecommand*{\iu}%
{\ensuremath{\mathrm{i}\,}}
\providecommand*{\eu}%
{\ensuremath{\mathrm{e}}}
\def\a{\alpha}
\def\b{\beta}
\def\c{\chi}
\def\d{\delta}
\def\D{\Delta}
\def\eps{\epsilon}
\def\vareps{\varepsilon}
\def\g{\gamma}
\def\G{\Gamma}
\def\grad{\nabla}
\newcommand{\h}{\frac{1}{2}}
\def\l{\lambda}
\def\om{\omega}
\def\s{\sigma}
\def\O{\Omega}
\def\p{\phi}
\def\vp{\varphi}
\def\P{\Phi}
\def\pa{\partial}
\def\pr{\prime}
\def\ti{\tilde}
\def\wti{\widetilde}
\newcommand{\cA}{\mathcal{A}}
\newcommand{\cB}{\mathcal{B}}
\newcommand{\cC}{\mathcal{C}}
\newcommand{\cD}{\mathcal{D}}
\newcommand{\cE}{\mathcal{E}}
\newcommand{\cF}{\mathcal{F}}
\newcommand{\cG}{\mathcal{G}}
\newcommand{\cH}{\mathcal{H}}
\newcommand{\cI}{\mathcal{I}}
\newcommand{\cJ}{\mathcal{J}}
\newcommand{\cK}{\mathcal{K}}
\newcommand{\cL}{\mathcal{L}}
\newcommand{\cM}{\mathcal{M}}
\newcommand{\cN}{\mathcal{N}}
\newcommand{\cO}{\mathcal{O}}
\newcommand{\cP}{\mathcal{P}}
\newcommand{\cQ}{\mathcal{Q}}
\newcommand{\cR}{\mathcal{R}}
\newcommand{\cS}{\mathcal{S}}
\newcommand{\cT}{\mathcal{T}}
\newcommand{\cU}{\mathcal{U}}
\newcommand{\cV}{\mathcal{V}}
\newcommand{\cW}{\mathcal{W}}
\newcommand{\cX}{\mathcal{X}}
\newcommand{\cY}{\mathcal{Y}}
\newcommand{\cZ}{\mathcal{Z}}
\newcommand{\nit}{\noindent}
\newcommand{\ct}[1]{\cite{#1}}
\newcommand{\bi}[1]{\bibitem{#1}}


\begin{center}
{\large\bf  Deformations of N=2 super-conformal algebra}
\end{center}
\begin{center}
{\large\bf  and supersymmetric two-component Camassa-Holm equation}
\end{center}
\normalsize
\vskip .4in

\begin{center}
 H. Aratyn

\par \vskip .1in \noindent
Department of Physics \\
University of Illinois at Chicago\\
845 W. Taylor St.\\
Chicago, Illinois 60607-7059\\
\par \vskip .3in

\end{center}

\begin{center}
J.F. Gomes and A.H. Zimerman

\par \vskip .1in \noindent
Instituto de F\'{\i}sica Te\'{o}rica-UNESP\\
Rua Pamplona 145\\
01405-900 S\~{a}o Paulo, Brazil
\par \vskip .3in

\end{center}

\begin{abstract}
This paper is concerned with a link between 
central extensions of $N=2$ superconformal
algebra and a supersymmetric two-component 
generalization of the Camassa--Holm equation.

Deformations of superconformal
algebra give rise to two compatible bracket structures.
One of the bracket structures is derived from the central 
extension and admits a momentum operator which agrees 
with the Sobolev norm of a coadjoint orbit element.
The momentum operator induces via Lenard relations 
a chain of conserved hamiltonians of the resulting 
supersymmetric Camassa-Holm hierarchy. 

\end{abstract}
\newpage
\section{Introduction}
\label{section:intro}
The last few years have greatly enhanced our
understanding of bihamiltonian structures.
Bihamiltonian structure is frequently viewed as one of the
key features of integrability and has recently been adopted as 
the basis for the classification program (see e.g. 
\ct{Magri78,Margi8,CaFaMaPe1,CaFaMaPe2,FaReZa,FaMaPe1,FaMaPe2,DuZh,
DuLiuZh,Loren})

In recent publications \ct{CH2,LiuZhang}, 
a study 
of deformations of bihamiltonian structures of hydrodynamic 
type led to a new class of 
bihamiltonian structure associated with  a two-component 
generalization of Camassa--Holm equation.
In this construction the second bracket component of the 
bihamiltonian structure contained a centerless Virasoro algebra 
\be
\{ U(x) , U(y) \}_2 = \d^{\pr} (x-y) \( U (x) +U (y) \)
= 2  U (x) \d^{\pr} (x-y) +  U^{\pr} (x) \d (x-y) 
\lab{clessvira}
\ee
coupled to a spin-one current via :
\be
\{ U(x) , v (y) \}_2 = \d^{\pr} (x-y) \( v (x) + (s-1) v (y) \), \quad
s=1\, .
\lab{spinoneira}
\ee
Two totally different hierarchies emerged  from deformations
of bihamiltonian system 
depending on whether a first or second bracket structure
has being deformed by addition of an appropriate central element.
Deformation of a second bracket led to a standard AKNS hierarchy 
while deformation of a first bracket gave rise to a new result:
the 2-component Camassa-Holm hierarchy.
The Hamiltonians of the bihamiltonian structure were shown to
split into two chains, one chain of the positive order inducing 
the AKNS hierarchy and another of the 
negative order inducing the 2-component Camassa-Holm 
hierarchy \ct{Aratyn:2005pg}.

Understanding of central extensions of a Poisson structure
is crucial for analysis of deformations.
Let us now consider a general Poisson bracket with a cocycle and 
link it to the question of compatibility
so essential in the theory of integrable systems. 
Generally for a given Poisson bracket structure
a cocycle of this bracket provides an additional compatible 
bracket structure \ct{magri}. A trivial cocycle 
routinely agrees with the conventional
first bracket structure. For example, shifting $v \to v+c$
in \rf{spinoneira}  leads  to a mixture of the second bracket
with the first bracket of the type 
$\{ U(x) , v (y) \}_1 = \d^{\pr} (x-y)$. 
A non-trivial cocycle can serve as a deformation of 
first or, equally well, second structure.
Algebraic properties of central elements ensure
compatibility of all three Poisson bracket structures, meaning 
compatibility of trivial cocycle, non-trivial cocycle and the 
original bracket structure.
In the setting of equation \rf{spinoneira} the non-trivial 
central element is proportional to $\d^{\pr \pr} (x-y)$ and it 
can be added to either \rf{spinoneira} or the first bracket 
$\{ U(x) , v (y) \}_1 = \d^{\pr} (x-y)$ (see equation \rf{bihamstr})
without violating compatibility 
of the underlying bihamiltonian structure.
The above arguments constitute an algebraic reasoning behind 
a trihamiltonian formalism introduced in \ct{tri-Hamiltonian}.

In this paper we will apply the above observations
to the N=2 superconformal algebra with a goal of building 
generalization of the 2-component Camassa-Holm model 
which is invariant under supersymmetric transformations.
This approach will not be based on N=2 superspace formalism, 
like the one used in \ct{popow2ch}, but instead it will
only employ the bihamiltonian method.

We will apply coadjoint orbit method to the $N=2$ 
superconformal model in order to develop a formalism, which 
derives equations of motion of the N=2 supersymmetric Camassa-Holm model
as Euler-Arnold equations.
Misiolek \ct{Misiolek}, showed that the Camassa-Holm equation can 
be characterized as a geodesic flow on the Bott-Virasoro group.
Guha and Olver \ct{GuhaOlver} obtained a fermionic (but not supersymmetric)
extension of Camassa-Holm equation as a geodesic flow.
Here we will fill this gap by providing a 
Sobolev $H^{(1)}$ inner-product which is 
compatible with supersymmetric structure
inherent in the $N=2$ superconformal algebra
and we will show that in this setting 
the supersymmetric Camassa-Holm equation 
follows as the Euler-Arnold equation.

Our formalism provides a direct guide on how to extend
the conventional $L_2$ inner product to its deformed 
version given by the Sobolev $H^{(1)}$ inner-product.
In all models  we consider the basic algebra defining the model
is inducing 
the second bracket structure. The compatible first bracket structure 
is then obtained by a central extension containing the deformation
parameter in front of the non-trivial cocycle.
In this setup the Sobolev $H^{(1)}$ inner-product is derived by demanding
that the norm of the generic algebra element is a momentum
of the first bracket structure. This norm is then used to
produce a chain of Hamiltonians defining the hierarchy
via Lenard relations.

This paper is organized as follows: In section \ref{section:2CH},
we review the construction of the two-component
Camassa-Holm equation within the coadjoint orbit formalism
emphasizing a general
link between deformation of the first bracket and the Sobolev norm
defining its momentum operator.
In section \ref{section:N2algebra}, we construct bihamiltonian
structure based on $N=2$ superconformal algebra with deformation
contained in the first bracket structure. The method developed
in section \ref{section:2CH} leads here to the momentum operator
defining generalized two-component Camassa-Holm hierarchy invariant 
under the $N=2$ supersymmetry transformations.
In section \ref{section:2coad}, we derive the $N=2$ supersymmetric 
Camassa-Holm equations as Euler-Arnold equations. 
In section \ref{section:hodographic}, we show that
the invariance under $N=2$ supersymmetry transformations 
extends to the $N=2$ Camassa-Holm model expressed in the hodographic
variables when combined with redefinition of the supersymmetry
transformations to ensure a closure of the $N=2$ algebra
in new variables.
We conclude the paper by presenting a brief outlook in 
section \ref{section:outlook}.

\section{Two-component Camassa-Holm model}
\label{section:2CH}
The semi-direct product $\cG$ of centerless Virasoro (Witt) and 
spin-one current algebra is given by
\be
\bigl[ (f,a),(g,b) \bigr]= \biggl(f g^{\pr} -f^{\pr} g, f b^{\pr} -a^{\pr}
g\biggr)
\lab{favira}
\ee
The brackets \rf{clessvira} and 
\rf{spinoneira} can then be derived using the Poisson bracket 
structure
on a coadjoint orbit $\cG^{*}$ induced by algebra commutation relations
via :
\be
\{ F, G \}_P \, (\mu) = \biggl\langle \mu \bigg| \biggl[ \frac{\d F}{\d \mu} ,
\frac{\d G}{\d \mu} \biggr]  \biggr\rangle\, ,
\lab{coad1}
\ee
where $F,G$ are real functions on $\cG^{*}$,  $\mu$ is in $\cG^{*}$
and $\langle \cdot | \cdot  \rangle$ is an inner-product.

Choosing $F$ as $F= \langle (U,v) | (f,a) \rangle$,
$G$ as $G= \langle (U,v) | (g,b) \rangle$ and 
$\mu=(U,v)$ in definition \rf{coad1} yields :
\be
\begin{split}
\biggl\{ \langle (U,v) \big| (f,a) \rangle \,,\, \langle (U,v) \big| (g,b) \rangle 
\biggr\}_P  
&= \biggl\langle (U,v)\, \big|\, \left[ (f,a),(g,b) \right]\biggr\rangle\\
&=\biggl\langle (U,v)\, \big|\, (f g^{\pr} -f^{\pr} g, f b^{\pr} -a^{\pr} g)
\biggr\rangle
\end{split}
\lab{3brack}
\ee
If $\langle \cdot | \cdot  \rangle$ is taken to be 
the $L_2$ inner-product :
\be 
\langle (U,v) | (f,a) \rangle_{L_2} = \int \left\lbrack
Uf + va \right\rbrack \rd x
\lab{L2}
\ee
then $\{ \cdot , \cdot \}_P$ in relation \rf{3brack} reproduces 
a structure given in \rf{clessvira} and 
\rf{spinoneira} together with 
\[\{ v(x) , v(y) \}_2 = 0\,. \]
The Lie algebra \rf{favira} is compatible with a central extension:
\be
\biggl[ (f,a,c_1,c_2),(g,b,c_1,c_2) \biggr]= 
\(0 , 0 , \om(a,b), \om (f, g)\)
\lab{virac}
\ee
using notation where $c_1$ and $c_2$ represent two additional
central directions and 
\[
\begin{split}
 \om(a,b)&= \h \langle (0,a^{\pr}), (0,b) \rangle =\h \int a^{\pr} b \rd x
, \\
 \om (f, g)&= \h \langle (f^{\pr},0), (g,0) \rangle
=\h \int \( f^{\pr} g+\b^2 f^{\pr\pr} g^{\pr} \) \rd x \, .
\end{split}
\]
The pairing \rf{L2} extends in a straightforward way to incorporate
two extra central elements and implies via relation \rf{3brack} the 
following new bracket structure:
\be
\begin{split}
\big\{ \langle (U,v) \big| (f,a) \rangle_{L_2} , 
\langle (U,v) \big| (g,b) \rangle_{L_2}
\big\}_1
&= \langle (U,v,c_1,c_2) \big| \(0 , 0 , \om(a,b), \om (f, g)\)
\rangle_{L_2}\\
&=
\h \int \big\lbrack c_1\( f^{\pr} g+\b^2 f^{\pr\pr} g^{\pr}\) +c_2\(a^{\pr}
b\) \big \rbrack \rd x \, .
\end{split}
\lab{2brack}
\ee
The above bracket is compatible with a second bracket $\{ \cdot, \cdot \}_2$
for all values of $c_1,c_2$.
For $c_1=c_2=1$ the above construction admits the following local 
first Poisson structure:
\be
\begin{split}
\{ U (x) , U(y) \}_1 &= \h \( 1-\b^2 \pa_x^2\) \d^{\pr} (x-y)  \\
\{ v(x) , U(y) \}_1 &= 0\\
\{ v(x) , v(y) \}_1 &= \h \d^{\pr} (x-y)\, .
\end{split}
\lab{U2brack-loc}
\ee
The existence of two compatible bracket structures signals
integrability and implies presence of infinitely many conserved 
Hamiltonian structures. To construct these Hamiltonians
we first search for the object which acts as a gradient 
with respect to the first bracket $\{ \cdot  , \cdot  \}_1$
from \rf{U2brack-loc}. We first investigate the $L_2$ norm of
$(U,v)$:
\[  \langle (U,v) \big| (U,v) \rangle_{L_2} =
 \int \( U^2 +v^2\) \rd x\, .
\]
For $h = (1/2) \langle (U,v) \big| (U,v) \rangle_{L_2}$
we find:
\[ \{ h , U(x) \}_1 = \h \( 1-\b^2 \pa_x^2\)  U^{\pr} (x),\quad
\{ h , v(x) \}_1 = \h   v^{\pr}(x) \, .
\]
By redefining $h$ as follows
\[ h\; \to \; {\bar h} = \h 
\int \( U  \( 1-\b^2 \pa_x^2\)^{-1} U  +v^2\) \rd x
\]
we now obtain desired relations:
\[ \{ {\bar h} , U(x) \}_1 = \h  U^{\pr} (x),\quad
\{ h , v(x) \}_1 = \h   v^{\pr}(x)
\]
but for the price of introducing a non-locality in the definition
of ${\bar h}$. To remove this non-locality we need to 
redefine variable $U$ to 
\[ U \quad \longrightarrow \quad m =  U -\b^2 U^{\pr\pr}\]
Replacing $U$ by $m$ in ${\bar h}$ yields
\[
\begin{split}
{\bar h}&\; \longrightarrow \;  H=\h 
\int \( m  \( 1-\b^2 \pa_x^2\)^{-1} m  +v^2\) \rd x \\
&= 
 \h 
\int \( U m  +v^2\) \rd x =
\h \int \( U^2 +\b^2 (U^{\pr})^2+v^2\) \rd x
\end{split}
\]
where we assumed periodic boundary conditions.
The above Hamiltonian $H$ appears as a norm of $(U,v)$:
\be
H = \h \langle (U,v) \big| (U,v) \rangle 
\lab{ham-norm}
\ee
with respect to the $H^{(1)}$ inner product:
\be
\langle (U,v) \big| (f,a) \rangle= \int \left\lbrack
Uf + \b^2 U^{\pr} f^{\pr} +va \right\rbrack \rd x
\lab{sobolev}
\ee
When  the bilinear form $\langle \cdot | \cdot  \rangle$  in 
definition \rf{coad1} is given by the above $H^{(1)}$ inner product 
then $\{ \cdot , \cdot \}_P$ from relation 
\rf{3brack} reproduces a familiar second 
bracket structure:
\be
\begin{split}
\{ m(x) , m(y) \}_2 &= 2 m(x) \d^{\pr} (x-y) + m^{\pr} (x) \d (x-y) \\
\{ v(x) , m(y) \}_2 &= v (x) \d^{\pr} (x-y) + v^{\pr} (x) \d (x-y) \\
\{ v(x) , v(y) \}_2 &= 0
\end{split}
\lab{3brack-loc}
\ee
Similarly replacing the $L_2$ inner-product by the $H^1$ inner product
in relation \rf{2brack} yields for the first bracket structure:
\be
\begin{split}
\{ m (x) , m(y) \}_1 &= \h \( 1-\b^2 \pa_x^2\) \d^{\pr} (x-y)  \\
\{ v(x) , m(y) \}_1 &= 0\\
\{ v(x) , v(y) \}_1 &= \h \d^{\pr} (x-y)
\end{split}
\lab{2brack-loc}
\ee
In the remaining part of this section  we will use the
$H^{(1)}$ inner product as defined in \rf{sobolev}.
Thus the canonical bracket structure $\{ \cdot  , \cdot \}_P$
agrees with that given by \rf{3brack-loc}.

The coadjoint action is obtained from
\be
\begin{split}
&\big\langle {\rm ad}^*_{(f,a)} (g,b) \big| (h,c) \big\rangle
=
\big\langle (g,b) \big| \left[ (f,a),(h,c) \right] \big\rangle\\
&=
\int \left\lbrack  g (f h^{\pr} -f^{\pr} h) +\b^2 g^{\pr} ( f h^{\pr} -f^{\pr} h)^{\pr}
+b (f c^{\pr} -a^{\pr} h) \right\rbrack \rd x \, .
\end{split}
\lab{coad}
\ee
For brevity we introduce the following notation:
\[
{\rm ad}^*_{(f,a)} (g,b)= (B_1 , B_0  ) \, .
\]
Then the left hand side of eq. \rf{coad} equals
\[
\int \( B_1 h + \b^2 B_1^{\pr} h^{\pr} +B_0 c\) \rd x
= \int \( h (1- \b^2 \pa_x^2) B_1 +B_0c\) \rd x \, .
\]
Comparing with the right hand side of eq. \rf{coad} we find
\be
\begin{split}
(1- \b^2 \pa_x^2) B_1 &= -f (1- \b^2 \pa_x^2) g^{\pr} -2 
f^{\pr} (1- \b^2 \pa_x^2) g -b a^{\pr}\\
B_0 &= -(bf)^{\pr}
\end{split}
\lab{b1b0}
\ee
Defining Euler equations as
\be
{{\pa}\over {\pa t}}(U,v) = - {\rm ad}^*_{(U,v)} (U,v)
\lab{euler}
\ee
we find from eq. \rf{b1b0} setting $f=g=U$, $a=b=v$ that
\be
{{\pa}\over {\pa t}}(U,v) = - \biggl( (1- \b^2 \pa_x^2)^{-1} \biggl[  
-U (1- \b^2 \pa_x^2) U^{\pr} -2 
U^{\pr} (1- \b^2 \pa_x^2) U -v v^{\pr}\biggr] , -(vU)^{\pr} \biggr)
\lab{euleru}
\ee
from which we derive the 2-component Camassa-Holm equations:
\be
\begin{split}
{{\pa}\over {\pa t}} (1- \b^2 \pa_x^2) U &=  3 U U^{\pr}  
-2 \b^2  U^{\pr} U^{\pr\pr}  -\b^2 U U^{\pr\pr\pr} +v v^{\pr} 
\\
{{\pa}\over {\pa t}} v &= (vU)^{\pr}
\end{split}
\lab{euleruv}
\ee
With the Hamiltonian $H$ from definition \rf{ham-norm}
and the bracket $\{ \cdot , \cdot \}_2$ from eq. \rf{3brack-loc},
the Euler equations are derived as Hamiltonian equations through:
\be
{{\pa}\over {\pa t}} (m,v)= \biggl( \{ m(x) , H \}_2, \{ v(x) , H \}_2 \biggr)=
\biggl( 2 m U^{\pr} + m^{\pr} U +v v^{\pr} , (vU)^{\pr}\biggr)
\lab{euler-ham}
\ee

Note, that the last of equations in \rf{2brack-loc} can also 
be written as
\be
\{ v^2 (x) , v^2(y) \}_1 = 2 v^2 (x) \d^{\pr} (x-y) + (v^2)^{\pr} (x) 
\d (x-y) \, .
\lab{2pvvsq}
\ee
With the above in mind we can rewrite two compatible bracket 
structures $\{ \cdot, \cdot \}_i, \, i=1,2$ in the matrix form
as follows:
\be
\begin{pmatrix} 
\{ m(x) , m(y) \}_i & \{ m(x) , v^2 (y) \}_i\\
\{ v^2(x) , m(y) \}_i & \{ v^2(x) , v^2 (y) \}_i \end{pmatrix}
= P_i \d (x-y), \qquad i=1, 2 \, .
\lab{pidef}
\ee
In the above basis we obtain
\be
P_2 = \begin{pmatrix} 
 m \pa +\pa m  & v^2 \pa +\pa  v^2 \\
v^2 \pa +\pa  v^2  & 0 \end{pmatrix}, \qquad
P_1 = \begin{pmatrix} 
\h \(1-\b^2 \pa^2\)\pa  & 0 \\
0  &  v^2 \pa +\pa  v^2 \end{pmatrix} \, ,
\lab{p3p2}
\ee
which give rise to the Hamiltonian operators.
Remarkably, one  now can define define a third local compatible
bracket structure through the formula $P_0= P_1 P_2^{-1} P_1$.
$P_0$ is  explicitly given by
\be
P_0 =  \begin{pmatrix} 
0& \h \(1-\b^2 \pa^2\)\pa   \\
\h \(1-\b^2 \pa^2\)\pa &  -m  \pa -\pa m  \end{pmatrix}
\lab{p1matr}
\ee
or 
\be
\begin{split}
\{ m(x) , m(y) \}_0 &= 0  \\
\{ v^2(x) , m(y) \}_0 &= \h \( 1-\b^2 \pa_x^2\) \d^{\pr} (x-y)\\
\{ v^2 (x) , v^2 (y) \}_0 &= -  
2 m (x) \d^{\pr} (x-y)- m^{\pr} (x) 
\d (x-y)  
\end{split}
\lab{1brack-loc}
\ee
Defining the recursion operator $R=P_1 P_0^{-1}$
we can write recurrence relation 
$P_2=R  P_1$, which connects the first and second 
bracket structure.

It is easy to see that the Hamiltonian $H$ from eq. \rf{ham-norm}
is a Casimir for the bracket structure $\{\cdot, \cdot\}_0$
given by \rf{1brack-loc}. We also note, that
$H$ acts as a gradient operator, 
$\{ X(x) , H\}_1= X^{\pr} (x)/2 $ for $X=m,v$,
for the first bracket 
structure, $\{\cdot, \cdot\}_1$, 
from \rf{2brack-loc}.

For new variables $w_1,w_2$ connected to $m,v$ via
\[
m= \h \(w_1 + \b w_1^{\pr}\), \quad w_2= v^2 + \frac{w_1^2}{4}
\]
the two lowest  Poisson brackets :
\be
\begin{split}
\{w_1(x), w_1(y)\}_1&=2 \delta'(x-y),\\
\{w_1(x), w_2(y)\}_1&=w_1(x) \delta'(x-y)+w_1^{\pr} (x) \delta(x-y),\\
\{w_2(x), w_2(y)\}_1&= 2 w_2(x)\d^{\pr} (x-y) +w_2^{\pr} (x) \delta(x-y)\, ,\\
\{w_1(x), w_1(y)\}_0&=\{w_2(x), w_2(y)\}_0=0,\\
\{w_1(x), w_2(y)\}_0&=\delta'(x-y)-\b \delta^{\pr\pr}(x-y).
\end{split}
\lab{bihamstr}
\ee
form for $\b=0$ a standard bihamiltonian structure which contains
the second bracket structure from  equations \rf{clessvira}
and \rf{spinoneira}.
For  a non-zero value of the deformation parameter, $\b\ne 0$, 
relations \rf{bihamstr}
describe a deformation of a bihamiltonian structure 
with the deformed first Poisson structure \ct{LiuZhang}.
This is that deformation of the bihamiltonian structure which 
gave rise to the two-component
Camassa-Holm equation in \ct{LiuZhang,CH2}.
In the next section we will focus on the similar deformation
of the larger $N=2$ superconformal Poisson algebra.

\section{N=2 Superconformal Algebra}
\label{section:N2algebra}
A starting point is a general second bracket structure (see for instance \cite{das} for $\epsilon = 1$ and $\kappa = 0$):
\be
\begin{split}
\{w_1(x), w_1(y)\}_2&=2 \eps \d^{\pr} (x-y),\\
\{w_1(x), w_2(y)\}_2&=w_1(x) \delta^{\pr} (x-y)+w_1^{\pr} (x) \delta(x-y)-\eps 
\d^{\pr\pr} (x-y) +\kappa \d^{\pr} (x-y),\\
\{w_2(x), w_2(y)\}_2&= 2 w_2(x)\d^{\pr} (x-y) +w_2^{\pr} (x) 
\delta(x-y)\, ,\\
\{w_1(x), \xi_1(y)\}_2&= \xi_1 (x) \d (x-y),\\
\{w_1(x), \xi_2(y)\}_2&= -\xi_2 (x) \d (x-y),\\
\{w_2(x), \xi_1 (y)\}_2&= 2 \xi_1 (x)\d^{\pr} (x-y) +\xi_1^{\pr} (x) 
\delta(x-y)\, ,\\
\{w_2(x), \xi_2 (y)\}_2&= \xi_2 (x)\d^{\pr} (x-y) \, ,\\
\{\xi_1(x), \xi_2 (y)\}_2&= - \frac{1}{4}
w_2 (x)\d (x-y) -\frac{1}{4} w_1 (x) \delta^{\pr} (x-y)-\frac{1}{4}
\eps \d^{\pr\pr} (x-y) -\frac{1}{4} \kappa \d^{\pr} (x-y),
\end{split}
\lab{n2secbra}
\ee
It distinguishes itself from the second bracket from \rf{clessvira},
\rf{spinoneira} by presence of additional 
fermionic variables $\xi_i,\, i=1,2$ and central elements.
The terms with $\eps$ describe deformation of a second bracket which is
compatible with Jacobi relations. The terms with a  constant $\kappa$ 
are trivial cocycle terms which can easily be absorbed by a shift
$w_1\to w_1-\kappa$. 

The algebra \rf{n2secbra} (with $ \kappa = 0$) can  easily be put in the form  considered 
by Labelle and Mathieu  \ct{Labelle} 
in terms of the following change of variables :
\be
\begin{split}
u_{LM} (x) & =  2w_2 (x) - w_1^{\pr}(x), \quad w_{LM} (x)  = i w_1 (x)\\
\xi_{LM1} (x) & = 2i\( \xi_1 (x) +\xi_2 (x) \)  , \quad 
\xi_{LM2} (x)  = 2\( \xi_1 (x) -\xi_2 (x) \) 
\end{split}
\lab{mathieu-variables}
\ee

The above second bracket is compatible with the first bracket structure:
\be
\begin{split}
\{w_1(x), w_1(y)\}_1&= 2 \b \delta^{\pr} (x-y)\\
\{w_1(x), w_2(y)\}_1&=\delta^{\pr} (x-y)-\b \delta^{\pr\pr}(x-y)\\
\{\xi_1(x), \xi_2 (y)\}_1 &=-\frac{1}{4} \delta^{\pr} (x-y)-
\frac{\b}{4} \delta^{\pr\pr}(x-y) \, , 
\end{split}
\lab{n2firstbra}
\ee
where terms with a constant $\b$ describe deformation of a 
first bracket which is compatible with the second bracket in \rf{n2secbra}.

Comparing the bracket structures in \rf{n2secbra} and \rf{n2firstbra}
we notice that they have identical central elements for
$\eps=\b$ and $\kappa=1$.

Next, we define the quantity :
\be
H = \int \rd x \left\lbrack 
w_1 \(1 +\b \pa_x\)^{-1} w_2 - \b \( \(1 +\b \pa_x\)^{-1} w_2\)^2
+4 \xi_1  \(1 -\b \pa_x\)^{-1} \xi_2\right\rbrack \, ,
\lab{H2grad}
\ee
which acts as a gradient 
with respect to the first bracket $\{ \cdot  , \cdot  \}_1$
from \rf{n2firstbra}, meaning that
\[ \{ X(x) , H\}_1= X^{\pr} (x)\quad {\rm for} \quad X=w_i, \xi_i,
\;\; i=1,2\,.
\]
The gradient operator $H$ contains non-localities and as in section
\ref{section:2CH} we will absorb non-localities within 
new redefined variables.
For this purpose we now introduce variables 
$\p_i, \eta_i, \; i=1,2$ 
such that they allow rewriting the first bracket \rf{n2firstbra}
as an Heisenberg-like algebra :
\be
\{w_i (x), \p_i (y)\}_1 =0 , \;\;
\{w_i (x), \p_j (y)\}_1 =\d^{\pr} (x-y) , \; i \ne j ,\;  i, j=1,2
\lab{wphi}
\ee
and
\[
\{\xi_2 (x), \eta_1 (y)\}_1 =
- \{\xi_1 (x), \eta_2 (y)\}_1 =
\frac{1}{4} \d^{\pr} (x-y) 
\, .
\]
It is easy to verify that original variables 
$w_i, \xi_i, \, i=1,2$ must be related to
$\p_i, \eta_i, \; i=1,2$ through relations:
\[
w_1= 2 \b \p_2 + \p_1 - \b \p_1^{\pr}, \;
w_2=\p_2+ \b \p_2^{\pr}, \; \xi_1=\eta_1+\b \eta_1^{\pr},\;
\xi_2=\eta_2 -\b \eta_2^{\pr} \, .
\]
Now, set
\[
u =\h \p_1, \; v =\h w_1
\]
and define $m$ as a linear combination
\[m= c_1 w_2+c_2 w_1+c_3 w_1^{\pr} \, .
\]
The coefficients $c_1,c_2,c_3$ will be chosen in such a way as to ensure
that $m$ and $v$ are decoupled, with respect to the first bracket structure,
meaning that the first bracket of $m$ with $v$ 
vanishes. In view
of equation \rf{wphi} this requirement implies that $m$ should not contain $\p_2$
or in different words that $m$ should be given by
\[m= c_1 \(w_2-\frac{1}{2\b} w_1-\h  w_1^{\pr}\)
\]
In the following we set $c_1=-\b$ and accordingly use
the definition:
\[ m= u - \b^2 u_{xx} = \h\(w_1 +\b w_{1 \, x}\) -\b w_2;
\quad \b \p_2 = v-u+\b u_{x}
\]
Note, that the first bracket \rf{n2firstbra} reads in terms of 
$m$, $v$ and $\xi_i, \eta_i, \, i=1,2$ as
\be
\begin{split}
\{ m(x) , m(y) \}_1 &= - \h\b \( 1-\b^2 \pa_x^2\) \d^{\pr} (x-y)  \\
\{ v(x) , m(y) \}_1 &= 0\\
\{ v(x) , v(y) \}_1 &= \h\b \d_{x} (x-y)\\
\{\xi_i (x), \eta_j (y)\}_1 &=- \eps_{ij} \frac{1}{4} \d^{\pr} (x-y) 
,\;  i, j=1,2 \, ,
\end{split}
\lab{n2firstmv}
\ee
where $\eps_{12}=-\eps_{21}=1$ and zero otherwise.
Also, note that the first three equations agree 
with the first bracket structure in \rf{2brack-loc} up
to an overall factor of $\pm \b$.
The gradient operator \rf{H2grad} becomes in this notation
\be
\begin{split}
H &= 
\int \lb \p_2 w_1 -\b \p_2^2 -4 \eta_2 \xi_1\rb \rd x \\
&=
-\frac{1}{\b} \int \rd x \left\lbrack 
u m  - v^2 - 4 \b \xi_1  \eta_ 2\right\rbrack 
\lab{H2gradn}
\end{split}
\ee
and it holds that 
$\{ X(x) , H\}_1= X^{\pr} (x)$ for $X=m, v, \xi_i,
\;\; i=1,2\,$.
It is important to point out that $H$ is invariant under 
the following $N=2$ supersymmetry transformation:
\begin{alignat}{2}
\d \xi_1 &= \frac{\eps_1-\eps_2}{2\b} \(m - (1+\b \pa_x) v\)
, &\quad
\d \xi_2 &=\frac{\eps_1+\eps_2}{2\b} \( -m +(1+\b \pa_x) v\)
\nonu\\
\d u &= \eps_1 \( \eta_1+\eta_2\) + \eps_2 \( \eta_1-\eta_2\) 
, &\quad
\d v &=\eps_1 \( \xi_1+\xi_2\) + \eps_2 \( \xi_1-\xi_2\) 
\lab{susyfirst}
\end{alignat}
meaning that $\d H=0$. This invariance of $H$ will later lead 
to invariance of the whole integrable model to be defined below.

Also, below we will see that $H$ will agree with the Sobolev norm
appearing in the coadjoint treatment of the model.

In terms of $m$ and $v$ and for $\kappa=\eps/\b$, the second bracket \
\rf{n2secbra} takes the following form :
\be
\begin{split}
\{ m(x) , m(y) \}_2 &= -\b \(2 m(x) \d_{x} (x-y) + m_{x} (x) \d (x-y)\)
- \h\eps \( 1-\b^2 \pa_x^2\) \d^{\pr} (x-y) \\
\{ v(x) , m(y) \}_2 &= -\b \(v (x) \d_{x} (x-y) + v_{x} (x) \d (x-y)\) \\
\{ v(x) , v(y) \}_2 &= \h\eps \d_{x} (x-y)\\
\{v (x), \xi_1(y)\}_2&= \h \xi_1 (x) \d (x-y),\\
\{v(x), \xi_2(y)\}_2&= -\h \xi_2 (x) \d (x-y),\\
\{m(x), \xi_1 (y)\}_2&= \h \xi_1 (x) \d (x-y)- \frac{\b}{2} \(
3 \xi_1 (x)\d^{\pr} (x-y) +\xi_1^{\pr} (x)\delta(x-y)\) \, ,\\
\{m(x), \xi_2 (y)\}_2&= -\h \xi_2 (x) \d (x-y)- \frac{\b}{2}
\(3 \xi_2 (x)\d^{\pr} (x-y) +\xi_2^{\pr} (x)\delta(x-y)\) \, ,\\
\{\xi_1(x), \xi_2 (y)\}_2&=  \frac{m(x)-v(x)}{4\b}\delta(x-y)
-\frac{1}{4}\(v^{\pr} \d (x-y) +2 v(x)  \delta^{\pr} (x-y)\)\\
&-\frac{\eps}{\b}\(\frac{1}{4} \delta^{\pr} (x-y)
+\frac{\b}{4} \delta^{\pr\pr}(x-y)\)
\, ,
\end{split}
\lab{n2secmv}
\ee
We notice that a cocycle of the second bracket provides an 
additional compatible bracket structure. From now on
we set $\eps$ to zero (while maintaining $\b \ne 0$ in \rf{n2firstbra}) 
and describe the bihamiltonian
structure consisting of brackets $\{\cdot,\cdot\}_1$ and 
$\{\cdot,\cdot\}_2$, which gives rise to Lenard relations:
\begin{equation}
\{w_i(x), H_{-j}\}_2=j \{w_i(x), H_{-j-1}\}_1,\quad\;\;\; j=1,2,3,{\ldots} \, .
\lab{lenard}
\end{equation}
The Lenard relations \rf{lenard} can be used to recursively build an 
hierarchy of Hamiltonians.
Starting with the gradient operator $H_{-2}=H$ from \rf{H2gradn}
we obtain from relations  \rf{lenard} an expression for
$H_{-1}=\int w_2(x) \rd x$, which happens to be 
the Casimir of the first Poisson bracket.
Similarly, we also find an expression for Hamiltonian $H_{-3}$ :
\[
H_{-3} =\h \int\lb w_1\p_1\p_2 -\b \p_1 \p_2^2 -4 \b \p_2 \eta_2 \eta_1
- 4 \p_1 \(2 \eta_2 \eta_1 +\b \eta_2 \eta_1^{\pr} \)
\rb \rd x
\, ,\nonu
\]
as an application of Lenard 
relations \rf{lenard} (for $\eps=0$).

We will study the flow of the bihamiltonian hierarchy defined
by :
\begin{equation}\lab{hamflows}
\begin{split}
\pder{w_i}{t}&= \{w_i(x), H_{-3}\}_1= \h 
\{w_i(x), H_{-2}\}_2 \\
\pder{\xi_i}{t}&= \{\xi_i(x), H_{-3}\}_1= \h 
\{\xi_i(x), H_{-2}\}_2 
,\; i=1,2 \, .
\end{split}
\end{equation}
In this way we obtain:
\be
\begin{split}
w_{1 \,t}&= \h \( w_1 \p_1-4 \b \eta_2 \eta_1\)_x \\
w_{2 \,t}&= \h \( 2\p_1\p_2 +\b \p_2^2+\b\p_1 \p_{2 \,x}-4 
(2 \eta_2\eta_1 + \b \eta_2 \eta_{1 \,x})\)_x \\
\xi_{1 \,t}&= \h \( 2\eta_1\p_1 +\b \eta_{1 \,x}\p_1+\b \eta_1 \p_2\)_x \\
\xi_{2 \,t}&= \h \( \b\eta_2\p_2 +2 \eta_2\p_1-\b (\eta_2 \p_1)_{x} \)_x \\
\end{split}
\lab{n2eqs}
\end{equation}
Equations of motion \rf{n2eqs} become in terms of $m,u$ and $v$:
\be
\begin{split}
m_{t}&= 3 u u_x -\b^2 u u_{x x x} -2 \b^2 u_{x} u_{x x}-
v v_x + \( \b \eta_2 \eta_1 +\b \xi_2 \eta_1 +\b \eta_2\xi_1\)_x \\
v_{t}&= \( v u + \b \eta_1 \eta_2\)_x\\
\xi_{1 \,t}&= \h \( 3 \eta_1 u +2 \b \eta_{1\,x}u+\eta_1 (v+\b u_x)\)_x \\
\xi_{2 \,t}&= \h \(3  \eta_2 u -2 \b\eta_{2\,x}u +\eta_2\( v - \b u_x  \) \)_x \\
\end{split}
\lab{n2meqs}
\end{equation}
Let 
\[ \psi_1= \( \xi_1 -\xi_2\)\sqrt{2},\quad
\psi_2= \( \xi_1 +\xi_2\)\sqrt{2}
\]
Then the fermionic part of \rf{n2secmv} takes in terms of $\psi_i , i=1,2$ ,
the following form : 
\be
\begin{split}
\{v(x), \psi_1(y)\}_2&= \h \psi_2 (x) \d (x-y),\\
\{v (x), \psi_2(y)\}_2&= \h\psi_1 (x) \d (x-y),\\
\{m(x), \psi_1 (y)\}_2&= \h \psi_2 (x)\d (x-y) 
-\frac{\b}{2} \( \psi_1^{\pr}(x) \delta(x-y) +3 \psi_1 (x) \d^{\pr} (x-y)\)\, ,\\
\{m(x), \psi_2 (y)\}_2&= \h \psi_1 (x)\d (x-y) 
-\frac{\b}{2} \( \psi_2^{\pr}(x) \delta(x-y) +3 \psi_2 (x) \d^{\pr} (x-y)\)\, ,\\
\{\psi_1 (x), \psi_1 (y)\}_2&= -\{\psi_2 (x), \psi_2 (y)\}_2=
\frac{1}{\b} \(m(x)-v(x) \) \d (x-y)\\
\{\psi_1 (x), \psi_2 (y)\}_2&=- 2 v(x) \d^{\pr} (x-y) -
v^{\pr} (x) \d (x-y) 
\end{split}
\lab{n2secpsi}
\ee
One can remove the deformation parameter $\b$ from the Poisson brackets
\rf{n2secmv} and \rf{n2secpsi} by defining
\be
M= -\frac{1}{\b} \(m-v\)
\lab{defM}
\ee
in terms of which the brackets involving $m$ can be expressed as :
\be
\begin{split}
\{ M(x) , M(y) \}_2 &= 2 M  (x) \d_{x} (x-y) + M_{x} (x) \d (x-y) \\
\{ v(x) , M(y) \}_2 &= v (x) \d_{x} (x-y) + v_{x} (x) \d (x-y) \\
\{M(x), \psi_i (y)\}_2&= \frac{1}{2}  \psi_i^{\pr} (x)\delta(x-y) +\frac{3}{2} 
\psi_i (x) \d^{\pr} (x-y)\, ,\;\; i=1,2\\
\{\psi_1 (x), \psi_1 (y)\}_2&= -\{\psi_2 (x), \psi_2 (y)\}_2=
- M(x) \d (x-y)
\end{split}
\lab{n2secM}
\ee
Thus the deformation parameter $\b$ has been effectively moved from
Poisson brackets to the redefined function $M$.

Introducing 
\[
\g_1=\eta_1-\eta_2, \quad \g_2=\eta_1+\eta_2
\]
we can rewrite $\psi_1,\psi_2$ as
\[ \psi_1= \( \g_1 +\b \g_{2\, x}\)\sqrt{2} ,\quad
\psi_2= \( \g_2 +\b \g_{1\, x}\)\sqrt{2} \, .
\]
In this notation the equations of motion \rf{n2meqs} become
\be
\begin{split}
m_{t}&= 3 u u_x -\b^2 u u_{x x x} -2 \b^2 u_{x} u_{x x}-
v v_x + \h  \( 3 \b \g_2 \g_1 +\b^2 \g_2 \g_{2\,x} -\b^2 \g_1\g_{1\,x}\)_x \\
v_{t}&= \( v u + \h \b \g_1 \g_2\)_x\\
\(\g_1 +\b \g_{2\,x}\)_{t}&= \h \( 3 u \g_1 +2 \b u \g_{2\,x}+ v \g_1 +\b u_x
\g_2\)_x \\
\(\g_2 +\b \g_{1\,x}\)_{t}&= \h \( 3 u \g_2 +2 \b u \g_{1\,x}+ v \g_2 +\b u_x
\g_1\)_x \, .
\end{split}
\lab{n2mgeqs}
\end{equation}
These equations of motions can be cast in a manifestly supersymmetric form
\be
\biggl[ \b \( D_1 D_2 \P \)-\P\biggr]_t=\biggl[ \b \P \( D_1 D_2 \P \)
-\P^2-\frac{\b}{2} \( D_2 \P \)\( D_1  \P \)\biggr]_x\, ,
\lab{popeq}
\ee
where the covariant derivatives $D_1,D_2$ defined 
in terms of Grassmannian variables $\theta_1, \theta_2$
as
\be
D_1=\pder{}{\theta_1}-\theta_1 \pa_x, \quad D_2=\pder{}{\theta_2}+\theta_2 \pa_x
\lab{d1d2}
\ee
satisfy
\[
D_1^2=-\pder{}{x} , \quad D_2^2=\pder{}{x}\, .
\]
and where we assumed that 
$\theta_i \gamma_j = -\gamma_j \theta_i$ for $i,j=1,2$.

The superfield 
\be
\P = u + \theta_1 \g_2 +\theta_2 \g_1 + \theta_2 \theta_1 \, \frac{u-v}{\b}
\lab{superf}
\ee
transforms under supersymmetry as 
\be
\d \P= \(\eps_1 Q_1 +\eps_2 Q_2 \) \P \, ,
\lab{susy}
\ee
where the generators $Q_1, Q_2$ are
\be
Q_1=\pder{}{\theta_1}+\theta_1 \pa_x, \quad Q_2=\pder{}{\theta_2}-\theta_2
\pa_x \, .
\lab{q1q2}
\ee
The supersymmetry transformation \rf{susy}
reads in components:
\be
\begin{split}
\d \g_1 &= \eps_1 \, \frac{u-v}{\b} +\eps_2 u_x \\
\d \g_2 &= -\eps_1 u_x -\eps_2 \, \frac{u-v}{\b} \\
\d u &= \eps_1 \g_2 +\eps_2 \g_1\\
\d v &= \eps_1 \( \g_2 +\b \g_{1\, x}\)+
\eps_2\( \g_1 +\b \g_{2\, x}\) \, .
\end{split}
\lab{susycomp}
\ee
It is easy to explicitly verify that the equations of motion \rf{n2mgeqs}
are invariant under supersymmetry transformations \rf{susycomp}.

Equation \rf{popeq} agrees with supersymmetric $N = 2,  \a= 4$
Camassa-­Holm equation obtained in \ct{popow2ch} using the hereditary recursion operator 
defined on a superspace.

\section{Coadjoint orbit method for ${\mathbf N=2}$ model}
\label{section:2coad}
Here we will develop a formalism, which derives equations
of motion \rf{n2meqs} as Euler-Arnold equations. 
The algebra elements of $N=2$ superconformal algebra 
$\cG$ will be denoted as $\( f, a , \a_1,\a_2\)$,
$\(g,b, \b_1,\b_2\)$ and so on.
They satisfy the commutation relations: 
\be
\begin{split}
&\biggl[ \( f, a , \a_1,\a_2\)\,,\, \(g,b, \b_1,\b_2\) \biggr]= 
\biggl( f g^{\pr} -f^{\pr} g -\a_2\b_1-\a_1 \b_2, \biggr. \\
&\biggl. f b^{\pr} -a^{\pr}g-\a_2\b_1-\a_1 \b_2-\b \a_2 \b_1^{\pr} +
\b \a_2^{\pr} \b_1 +\b \a_1 \b_2^{\pr} - \b \a_1^{\pr} \b_2,  \biggr. \\
&\biggl. -\frac{1}{2 \b} a \b_1 +\frac{1}{2 \b} b \a_1 +\frac{1}{2 \b} f \b_1
-\frac{1}{2 \b} g \a_1+f \b_1^{\pr} -\h f^{\pr} \b_1 - g \a_1^{\pr} 
+ \h g^{\pr} \a_1, \biggr.\\
&\biggl. \frac{1}{2 \b} a \b_2 -\frac{1}{2 \b} b \a_2 -\frac{1}{2 \b} f \b_2
+\frac{1}{2 \b} g \a_2+f \b_2^{\pr} -\h f^{\pr} \b_2 - g \a_2^{\pr} 
+ \h g^{\pr} \a_2 \biggr)
\end{split}
\lab{n2viraa}
\ee
Its dual space will be denoted as $\cG^*$. The typical element
of $\cG^*$ is denoted by $\(u,v, \eta_1,\eta_2\)$
The pairing between $\cG$ and $\cG^*$ will be provided by 
$H^{(1)}$ inner-product defined as:
\be
\begin{split}
\langle (u,v,\eta_1,\eta_2) \big| (f,a,\a_1,\a_2) \rangle&= 
\int \bigl[ \frac{va}{\b} -\frac{1}{\b} \( u f + \b^2 u^{\pr} f^{\pr} \)
+2 \( \eta_1+ \b \eta_1^{\pr} \) \a_2-
2 \( \eta_2- \b \eta_2^{\pr} \) \a_1\bigr] \rd x \\
&= \int \bigl[ \frac{va}{\b} -\frac{1}{\b} f m +2 \xi_1 \a_2-
2 \xi_2\a_1\bigr] \rd x
\lab{soboln=2}
\end{split}
\ee
The coadjoint action is obtained from
\be
\begin{split}
\langle {\rm ad}^*_{(f,a,\a_1,\a_2)} (u,v,\eta_1,\eta_2) \big| (g,b,\b_1,\b_2)
\rangle &=
\langle (u,v,\eta_1,\eta_2) \big| \left[ \( f, a , \a_1,\a_2\)\,,\, \(g,b, \b_1,\b_1\) \right] \rangle\\
\end{split}
\lab{coadn2}
\ee
Denote 
${\rm ad}^*_{(f,a,\a_1,\a_2)} (u,v,\eta_1,\eta_2)$
by $\(B,B_0,B_1, B_2\)$. Then equation \rf{coadn2} yields
\[
\begin{split}
\(1-\b^2 \pa_x^2\) B&=  - (f m)^{\pr} -f^{\pr} m
+ v a^{\pr} - (\xi_1 \a_2 +\xi_2 \a_1)
+2\b \( \xi_1 \a_2^{\pr}-\xi_2\a_1^{\pr}\)
+\b \( \xi_1 \a_2-\xi_2\a_1\)^{\pr}\\
B_0 &= -\( fv\)^{\pr} +\xi_1\a_2+\xi_2\a_1\\
\(1+\b \pa_x\) B_{1}&= \h \(\frac{1}{\b}\a_1 m-\a_1^{\pr} v -
\(\a_1 v\)^{\pr}-\frac{1}{\b} \a_1 v +\frac{1}{\b} \xi_1 a
-\frac{1}{\b} \xi_1f-\xi_1 f^{\pr} -2\(\xi_1 f\)^{\pr}\)\\
\(1-\b \pa_x\) B_{2}&= -\h \(\frac{1}{\b}\a_2 m+\a_2^{\pr} v +
\(\a_2 v\)^{\pr}-\frac{1}{\b} \a_2 v +\frac{1}{\b} \xi_2 a
-\frac{1}{\b} \xi_2 f+\xi_2 f^{\pr}+ 2\(\xi_2 f\)^{\pr}\)
\end{split}
\]
Substituting $(u,v,\eta_1,\eta_2)$ for $(f,a,\a_1,\a_2)$ in the above
equations we  see that equations
of motion \rf{n2meqs} are reproduced as the Euler-Arnold equations: 
\be
\dder{}{t} (u,v,\eta_1,\eta_2) = - {\rm ad}^*_{(u,v,\eta_1,\eta_2)} (u,v,\eta_1,\eta_2)
\lab{eulern2}
\ee
The above equation can also be derived using the Poisson bracket structure
on a coadjoint orbit $\cG^{*}$ induced by algebra commutation relations
via relation \rf{coad1}.
For 
\be
F= \langle (u,v,\eta_1,\eta_2)  | (f,a,\a_1,\a_2) \rangle, \;\; 
G= \langle (u,v,\eta_1,\eta_2)  | (g,b,\b_1,\b_2) \rangle, \;\; 
\mu=(u,v,\eta_1,\eta_2) 
\lab{FGmu}
\ee
we find from relation \rf{coad1} 
the bracket structure which agrees with $\{ \cdot , \cdot \}_2$ from
\rf{n2secmv}.  

After multiplying the right hand side of equation \rf{eulern2} by 
$| (f,a,\a_1,\a_2) \rangle$ we obtain
\[
 - \langle {\rm ad}^*_{(u,v,\eta_1,\eta_2)} (u,v,\eta_1,\eta_2)
 | (f,a,\a_1,\a_2) \rangle = 
 \langle  (u,v,\eta_1,\eta_2)
 | {\rm ad}_{(f,a,\a_1,\a_2)} (u,v,\eta_1,\eta_2) \rangle
\]
In view of definition \rf{coad1} and due to the above relation 
the Euler-Arnold equation can be cast in a form:
\[
\dder{}{t} F= \{ F, H\}
\]
for $F$ given by definition \rf{FGmu} and for $H$ being a 
half of the $H^1$-like norm of $(u,v,\eta_1,\eta_2)$ given 
explicitly by
\be
\begin{split}
H&=\h \langle (u,v,\eta_1,\eta_2) \big| (u,v,\eta_1,\eta_2) \rangle\\
&= 
\h \int \bigl[ \frac{v^2}{\b} -\frac{1}{\b} \( u^2 + \b^2 (u^{\pr})^2 \)
+2 \( \eta_1+ \b \eta_1^{\pr} \) \eta_2-
2 \( \eta_2- \b \eta_2^{\pr} \) \eta_1\bigr] \rd x \\
&= \h H_{-2}
\lab{hamnormn=2}
\end{split}
\ee
with $H_{-2}$ as defined in \rf{H2gradn}.

A central extension of algebra \rf{n2viraa} is given by
the following commutation relations: 
\be
\begin{split}
&\biggl[ \( f, a , \a_1,\a_2,c,c_0,c_{12}\)\,,\, \(g,b, \b_1,\b_2,c,c_0,c_{12}
\) \biggr]_c\\
&=\frac{\eps}{2\b^2} \(0,0,0,0,\int
f(1-\b^2 \pa_x^2)g^{\pr},\int a b^{\pr} ,  {2\b} \int (1-\b \pa_x)\a_2^{\pr}\b_1-
(1+\b \pa_x)\a_1^{\pr}\b_2 \)
\end{split}
\lab{n2virac}
\ee
Via extension of relation \rf{coad1} the above central extension leads to
the Poisson bracket structure
\be
\begin{split}
\{ m(x) , m(y) \}_c &= - \h\eps \( 1-\b^2 \pa_x^2\) \d_{x} (x-y)  \\
\{ v(x) , m(y) \}_c &= 0\\
\{ v(x) , v(y) \}_c &= \h\eps \d_{x} (x-y)\\
\{\xi_1(x), \xi_2 (y)\}_c &=\frac{\eps}{\b}\(-\frac{1}{4} \delta^{\pr} (x-y)-
\frac{\b}{4} \delta^{\pr\pr}(x-y)\)
\end{split}
\lab{n2cpb}
\ee
which agrees both with the first bracket structure in \rf{n2firstmv}
and the central extension of the second bracket structure in \rf{n2secbra}.

Note, that the quantity
\[
H_{-1} =\frac{1}{\b} \int (v-m) \rd x
\]
commutes with all variables with respect to the first bracket 
structures \rf{n2cpb} or \rf{n2firstmv}. 
With respect to the second bracket structure \rf{n2secbra}
$H_{-1}$ acts as a gradient operator :
\be
\{ X(x), H_{-1} \}_2 = X^{\pr} (x) , \quad {\rm for}\;\;
X=m,v,\xi_1,\xi_2
\lab{Hgrad}
\ee
We find from the above that 
\[ 
\begin{split}
\{ F, H_{-1} \}_2&= \langle (u^{\pr},v^{\pr},\eta_1^{\pr},\eta_2^{\pr})  
\bigg| (f,a,\a_1,\a_2) \rangle\\
&=\int \bigl[ \frac{v^{\pr} a}{\b} -\frac{1}{\b} f \( 1- \b^2 \pa_x^2\)
u^{\pr}+2 \( 1+ \b \pa_x\)\eta_1^{\pr}  \a_2-
2 \( 1- \b \pa_x \)\eta_2^{\pr} \a_1\bigr] \rd x \\ 
&= \frac{\eps}{2 \b}
\biggl\langle (u,v,\eta_1,\eta_2)  
\bigg|
\biggl[ \( f, a , \a_1,\a_2,c,c_0,c_{12}\)\,,\, \(u,v, \eta_1,\eta_2,c,c_0,c_{12}
\) \biggr]_c\biggr\rangle
\end{split}
\]
After setting $\eps=\b$ we obtain that the above equation
equals
\[
\begin{split}
\{ F, H_{-1} \}_2&=
\h \biggl\langle (u,v^{\pr},\eta_1^{\pr},\eta_2^{\pr})  
\bigg|
\biggl[ \( f, a , \a_1,\a_2,c,c_0,c_{12}\)\,,\, \frac{\d}{\d \mu} 
\v (u,v,\eta_1,\eta_2)  \v^2 \biggr]_c \biggr\rangle \\
&=\{ F, H \}
\end{split}
\]
for $H$ from \rf{hamnormn=2}, $F$ from \rf{FGmu} and with  
$\mu=(u,v,\eta_1,\eta_2) $.

Consider the first bracket structure given by \rf{n2cpb} with $\eps=\b$
or \rf{n2firstmv}, then the Hamiltonian $H$ from \rf{hamnormn=2}
serves as a momentum with respect to this bracket structure 
according to:
\be
\{ X(x), H \}_1 = \{ X (x), H_{-1} \}_2 =X^{\pr} (x) , \quad {\rm for}\;\;
X=m,v,\xi_1,\xi_2
\lab{Hmomentum}
\ee
as follows from \rf{Hgrad} and the Lenard relation proved  above.
This shows that the $H^1$-like norm is fully fixed by the requirement
that the norm $H$ serves as a momentum with respect to the
first bracket structure defined by a central extension.
Furthermore applying the Lenard argument we find
\[
\{ X(x), H_{-1} \}_1 = \{ X (x), H_{0} \}_2
\]
with $H_0$ being a constant by a power of fields counting  
argument. Thus, we have proved that $H_{-1}$ is a Casimir of 
the first bracket structure.

\section{Hodographic transformation and Supersymmetry}
\label{section:hodographic}
It is customary to study Camassa-Holm equation in terms 
of hodographic variables (see for instance \ct{LiuZhang}). The basic question which we will address in this section
is how the change of variables to
the hodographic variables affects supersymmetry described above.
The difficulty one encounters here is the fact that the 
 hodographic variables do not commute with the regular supersymmetry
 transformations. This obstacle makes it necessary to define
 a new  set of extended supersymmetry  transformations 
obeying $N=2$ supersymmetric algebra and commuting with hodographic 
variables.
In the next step we rewrite all equations of motion
\rf{n2mgeqs} as continuity equations in terms of the 
hodographic variables which leads to the proof 
of their invariance under the $N=2$ supersymmetry transformations.

Here, we will study a hodographic change of variables based on
a continuity equation, $v_{t}= \( v u + \h \b \g_1 \g_2\)_x$,
the second of equations in \rf{n2mgeqs}.
It follows from this continuity equation that a one-form: 
\[
\omega = v \rd x + \( vu  + \h \b \g_1 \g_2\) \rd t
\]
is closed, meaning that $ \rd \om=0$ with 
$\rd= \rd x \pder{}{x} + \rd t \pder{}{t}$.
Thus, it is possible to rewrite $\om$ as
\[ 
\om =\rd y
\]
for some scalar function $y$. Let $s$ be such that $\rd s = \rd t$
and choose now a new coordinate system with coordinates 
$y$ and $s$.
Then in the new coordinate system an expression for the
exterior derivative $\rd $ becomes
\[
\rd = \rd y \pder{}{y} + \rd s \pder{}{s}
= \(v \rd x + \( vu  + \h \b \g_1 \g_2\) \rd t\) 
\pder{}{y} + \rd t \pder{}{s} \, .
\]
Comparing with expression $\rd= \rd x \pder{}{x} + \rd t \pder{}{t}$
it follows that
\be
\pder{}{x} = v \pder{}{y}, \quad 
\pder{}{t} = \pder{}{s} + \( vu  + \h \b \g_1 \g_2\) \pder{}{y}
\lab{sy-der}
\ee
The fact that $\rd^2=0$ ensures commutativity 
\[
\left[ \pder{}{y} , \pder{}{s} \right]=0
\]
of derivatives with respect to the new variables.

{}From the above construction we get
\[
\rd x = - \( u  + \h \b \g_1 \g_2/v \) \rd s + 
\frac{1}{v} \rd y\, .
\]
Thus 
\be
\(\frac{1}{v}\)_s = - \( u  + \h \b \g_1 \g_2/v \)_y \, .
\lab{contis1}
\ee
{}From \rf{sy-der} we find that 
\[
 \pder{x}{y}=\frac{1}{v} , \qquad 
\pder{x}{s}=- \( u  + \h \b \g_1 \g_2/v \)\, .
\]
The first of equations \rf{n2mgeqs} can also be rewritten as a continuity
equation. Define namely 
\be
P= \frac{m}{v^2} + \b \frac{\psi_1 \psi_2 }{2 v^3}
\lab{def-P}
\ee
and 
\be
\rho=-v - \frac{\b}{2} P \g_1 \g_2 + \frac{1}{2v}
\( 2 \b \g_2 \g_1 +\b^2 \g_2 \g_{2\,x} -\b^2 \g_1\g_{1\,x}\)\, .
\lab{def-rho}
\ee
Then the first of equations \rf{n2mgeqs} amounts to:
\be P_s = \rho_y
\lab{contis2}
\ee
Thus, there exists a potential $f$ such that
\[ P=f_y , \qquad \rho =f_s\]
The last two equations of \rf{n2mgeqs} represent fermionic
continuity equations and can be used to define fermionic coordinates
$\tau_i, \, i=1,2$ through e.g.
\[ \rd \tau_1 = \frac{1}{\sqrt{2}} \psi_1 \rd x + 
\h \( 3 u \g_1 +2 \b u \g_{2\,x}+ v \g_1 +\b u_x
\g_2\) \rd t \, .
\]
The coordinate $\tau_1$ satisfies
\[
\pder{\tau_1}{x}=\frac{1}{\sqrt{2}} \psi_1 , \qquad
\pder{\tau_1}{t} = \h \( 3 u \g_1 +2 \b u \g_{2\,x}+ v \g_1 +\b u_x
\g_2\) \, .
\]
In hodographic variables the fermionic conservation laws become:
\be
\begin{split}
\(\frac{\psi_1}{\sqrt{2}v}\)_s&= \frac{1}{2}\(u \g_1+v\g_1+\b u_x \g_2
-\frac{\b}{v} \g_1\g_2 \frac{\psi_1}{\sqrt{2}}\)_y\\
\(\frac{\psi_2}{\sqrt{2}v}\)_s&= \frac{1}{2}\(u \g_2+v\g_2+\b u_x \g_1
-\frac{\b}{v} \g_1\g_2 \frac{\psi_2}{\sqrt{2}}\)_y \, .
\end{split}
\lab{contis3}
\ee
Thus, after the hodographic transformation all equations of motion 
of the model are described by continuity equations 
\rf{contis1}, \rf{contis2}, \rf{contis3}, which all have a common form:
\[
G_s =H_y \, .
\]
Now, we will investigate invariance of such continuity equations under
the supersymmetry transformations \rf{susycomp}.
Consider the $\eps_1$ part of the supersymmetry transformations \rf{susycomp}:
\begin{alignat}{2}
\frac{1}{\sqrt{2}} \d_1  \psi_1 &= \eps_1 \, \frac{m-v}{\b},& \qquad
\frac{1}{\sqrt{2}} \d_1  \psi_2 &= -\eps_1 \, v_x \nonu\\
\d_1 v &= \eps_1 \frac{1}{\sqrt{2}} \psi_2  ,& \qquad
\d_1 m &= \eps_1 \( \frac{1}{\sqrt{2}} \psi_2- \frac{\b}{\sqrt{2}} 
\psi_{1\,x}\)
\nonu
\end{alignat}
For an arbitrary function $F (y,s)$ it holds on basis of \rf{sy-der} 
that 
\be
\begin{split}
\d_1 \pder{}{y} F&=\pder{}{y} \d_1 F - \eps_1 \frac{\g_2+\b \g_{1\,x}}{v}
\pder{}{y} F\\
\d_1 \pder{}{s} F&=\pder{}{s} \d_1 F - \frac{\eps_1}{2}
\(u \g_2+v\g_2+\b u_x \g_1
-\frac{\b}{v} \g_1\g_2 \frac{\psi_2}{\sqrt{2}}\)
\pder{}{y} F
\end{split}
\lab{d1-noncom}
\ee
Let $h_1$ be a function  such that
\[
h_{1\,y} =\frac{\psi_2}{\sqrt{2}v}, \quad
h_{1\,s}= \frac{1}{2}\(u \g_2+v\g_2+\b u_x \g_1
-\frac{\b}{v} \g_1\g_2 \frac{\psi_2}{\sqrt{2}}\)
\]
in agreement with \rf{contis3}.
Applying $\d_1$ on a general continuity equation
$G_s =H_y$ yields
\[
\( \d_1 G - \eps_1 h_1 G_y\)_s=
\( \d_1 H - \eps_1 h_1 G_s\)_y=
\( \d_1 H - \eps_1 h_1 H_y\)_y \, ,
\]
which can be rewritten as a new transformed continuity 
equation :
\[
\( {\wti \d}_1 G\)_s = \( {\wti \d}_1 H\)_y \, ,
\]
where we introduced an extended supersymmetry transformation
\[
 {\wti \d}_1 =\d_1-\eps_1 h_1 \pder{}{y} \, .
\]
The action of ${\wti \d}_1$ is explicitly given by
\be
\begin{split}
 {\wti \d}_1  \g_1 &= \eps_1 \, \frac{u-v}{\b} -\eps_1 h_1 \g_{1\,y}\\
 {\wti \d}_1  \g_2 &= -\eps_1 u_x -\eps_1 h_1 \g_{2\,y} \\
 {\wti \d}_1  u &= \eps_1 \g_2 -\eps_1 h_1 u_{y}\\
 {\wti \d}_1 v &= \eps_1 \( \g_2 +\b \g_{1\, x}\)
 -\eps_1 h_1 v_y \, .
\end{split}
\lab{esusy1}
\ee
Consider now the $\eps_2$ part of the supersymmetry 
transformations \rf{susycomp} and define 
$h_2$ such that
\[
h_{2\,y} =\frac{\psi_1}{\sqrt{2}v}, \quad
h_{2\,s}= \frac{1}{2}\(u \g_1+v\g_1+\b u_x \g_2
-\frac{\b}{v} \g_1\g_2 \frac{\psi_1}{\sqrt{2}}\)
\]
In this case the extended supersymmetry transformation
is defined as
\[
 {\wti \d}_2 =\d_2-\eps_2 h_2 \pder{}{y}
\]
and explicitly given by
\be
\begin{split}
 {\wti \d}_2  \g_1 &= \eps_2 \, u_x  -\eps_2 h_2 \g_{1\,y}\\
 {\wti \d}_2  \g_2 &= -\eps_2 \frac{u-v}{\b} -\eps_2 h_2 \g_{2\,y} \\
 {\wti \d}_2  u &= \eps_2 \g_1 -\eps_2 h_2 u_{y}\\
 {\wti \d}_2 v  &= \eps_2 \( \g_1 +\b \g_{2\, x}\)
 -\eps_2 h_2 v_y \, .
\end{split}
\lab{esusy2}
\ee
The supersymmetry transformation ${\wti \d}_2$
maps the continuity equation $G_s =H_y$ into
\[
\( {\wti \d}_2 G\)_s = \( {\wti \d}_2 H\)_y \, .
\]
The main point is that the continuity equations 
$\( {\wti \d}_i G\)_s = \( {\wti \d}_i H\)_y$ for $i=1,2$
are linear combinations of original continuity equations
\rf{contis1}, \rf{contis2},
\rf{contis3}. 
Thus, the extended supersymmetry transformations 
${\wti \d}_i, \; i=1,2$
keep the model invariant and we conclude that
\be
\lbrack {\wti \d}_i \, ,\, \pder{}{y} \rbrack= 0, \quad
\lbrack {\wti \d}_i \, ,\, \pder{}{s} \rbrack= 0\, .
\lab{extcommut}
\ee
To find the action of the extended supersymmetry 
transformations ${\wti \d}_i, \; i=1,2$ 
on $h_i, \; i=1,2$ we first calculate :
\[
{\wti \d}_1 h_{1 \,y}= -\eps_1 \(v + h_1h_{1 \,y}\)_y,
\quad
{\wti \d}_2 h_{2 \,y}= \eps_2 \(v - h_1h_{1 \,y}\)_y
\]
and
\[
{\wti \d}_2 h_{1 \,y}= -\eps_2 \(\frac{m-v}{\b v} + \(h_2  h_{1 \,y}\)_y\),
\quad
{\wti \d}_1 h_{2 \,y}= \eps_1 \(\frac{m-v}{\b v} - \(h_1h_{2 \,y}\)_y\)
\]
Thus,
\[
{\wti \d}_1 h_{1 }= -\eps_1 \(v + h_1h_{1 \,y}+c_{11}\),
\quad
{\wti \d}_2 h_{2 }= \eps_2 \(v - h_1h_{1 \,y}+c_{22}\)
\]
and
\[
{\wti \d}_2 h_{1 }= -\eps_2 \(\int \rd y \frac{m-v}{\b v} + h_2  h_{1
\,y}+c_{21}\),
\quad
{\wti \d}_1 h_{2 \,y}= \eps_1 \(\int \rd y \frac{m-v}{\b v} - h_1h_{2 \,y}
+c_{12}\)
\]
where $c_{ij}, i,j=1,2$ are integration constants.

Despite a presence of non-local terms in the above extended supersymmetry
transformations one finds that their repeated application yields 
\[
{\wti \d}_1^{\pr} {\wti \d}_1 X = \pm \eps_1\eps_1^{\pr} c_{11} X_y, \quad
{\wti \d}_2^{\pr} {\wti \d}_2 X = \pm \eps_2\eps_2^{\pr} c_{22} X_y
\] 
where $X$ stands for $u,v, \g_1, \g_2$.
Similarly,
\[ \({\wti \d}_2 {\wti \d}_1 -
{\wti \d}_1 {\wti \d}_2\) X =  \eps_1\eps_2 \(c_{21}-c_{12}\)  X_y
\]
Choosing the integration constants as $c_{11}=c_{22}=1$ and 
$c_{21}=c_{12}$ we reproduce 
the $N=2$ algebra of the supersymmetry transformations given in eq. 
\rf{susycomp}.

\section{Outlook}
\label{section:outlook}

In this paper we have described a general 
formalism which associates a bihamiltonian structure 
to the $N=2$ superconformal Poisson algebra by taking advantage
of existence of central extension of the original algebra.
We have shown how this deformation uniquely determines a
deformed Sobolev-type inner product for the coadjoint orbit 
formalism through construction of the momentum operator for 
the first bracket.
The momentum operator is invariant under the supersymmetry 
transformation and induces via Lenard relations 
a chain of conserved hamiltonians determinng the 
$N=2$ supersymmetric Camassa-Holm hierarchy. 

All the steps of the above construction are general 
in nature and the formalism extends easily to other
similar formalisms. 
In a forthcomming publication \ct{next} we will apply the formalism
to the $N=1$ superconformal algebra and also provide formal proofs for 
the observations made here.

\section*{\sf Acknowledgments}
H.A. acknowledges partial support from Fapesp and IFT-UNESP 
for their hospitality. JFG and AHZ thank CNPq for a partial support.
The authors thank Leandro H. Ymai for help and discussions

\end{document}